# Enhancing Marine Data Transmission with Socially-Aware Resilient Vessel Networks

Ruobing Jiang, *Member, IEEE*, Chao Liu, *Member, IEEE*

*Abstract*—With the multi-dimensional exploration towards oceans, enormous sensing data has been generated with significant volume, velocity, variety and heterogeneity. The resulted Big Marine Data (BMD) thus issue unprecedented architectural challenges on existing marine communication systems. Current dominant marine communication technologies, e.g., shore-based cellular stations, high frequency radio, and expensive satellites, extremely suffer from short coverage, low bandwidth, insecurity, and unavailable cross-domain transmission. In this paper, Resilient Vessel Network (RVN) is proposed to fundamentally enhance BMD transmission. RVNs with widespread self-organized vessels and opportunistic connections reveal advantages of ubiquity, resilience, low cost and cross-domain transmission. To efficiently manage opportunistic vessel-to-vessel (V2V) connections for optimal routing, Social Network Analysis (SNA) on historical vessel interactions is applied for vessel familiarity measurement and community detection. The performance of the proposed community-based routing (CBR) is comprehensively evaluated with real datasets of fishing vessel trajectories. It is demonstrated that CBR achieves much lower transmission cost with comparable delivery ratio compared to typical routing algorithms.

## I. INTRODUCTION

**B**ACKGROUND on Big Marine Data. With the comprehensive human efforts to recognize, monitor, exploit, and protect the oceans, vast amount of diverse marine data has been generated at an amazing velocity. The annual data volume managed by US National Oceanic and Atmospheric Administration (NOAA) reached 30 PB back in 2012 [1]. In recent years, the annual increment of marine data kept by the Chinese Ministry of Natural Resources has also reached PB level. As illustrated in **Fig. 1**, multi-dimensional marine data can be sensed by underwater, surface, or remote sensors, e.g., seabed sensors, underwater cameras, sea gliders, exploration platforms, buoys, and remote sensing satellites. The generated Big Marine Data (BMD) plays a vital role in a wide range of areas such as oceanographic research, marine ecological monitoring and protection, marine disaster forecasting and mitigation, submarine geomorphology observation, marine resource exploration, and international exchanges and cooperation. However, the efficient transmission of BMD will become a significant challenge when the original high volume, velocity, and variety of big data are coupled with the heterogeneity and complexity unique to marine data.

**Limitations of Existing Communication Systems**. Unfortunately, existing marine communication technologies cannot facilitate wide-coverage, low-cost, and stable BMD transmission. First, shore-based cellular base stations can only support offshore communications within 30 km of the nearshore. Second, though communication satellites have extensive coverage, the cost is significant with limited throughput. For example, the single-beam capacity of the Asia-Pacific 6D satellite, launched in 2020 with the largest communication capacity in China, is only 1 Gbps for the area of up to 600-850 km in diameter. Third, typical offshore wireless communication systems are unstable with low bandwidth, e.g., the IF-based NAVTEX navigation warning system, the HF-based PACTOR communication system, and the VHF-based automatic identification system (AIS). Finally, the underwater transmission involves the bottleneck of long and medium range transmission.

**Challenges of BMD Transmission.** There exist several major challenges for BMD transmission. 1) Sufficient spatial-temporal coverage. The only available communication beyond 30 km offshore is through the costly and low throughput satellites, requiring dedicated terminals. 2) High communication bandwidth. The great volume of BMD requires high bandwidth, which cannot be provided by existing communication systems. 3) Secure and reliable transmission. The national marine data of each country include seabed topography, oil and gas resources, biological and ecological environments, which are extremely sensitive and confidential. 4) Cross-domain transmission. Transmission barriers between different domains, e.g., underwater and surface sensing, are caused by diverse communication devices, protocols and cross-media propagation capabilities of different signals.

**Our Approach.** We propose to use Resilient Vessel Networks (RVNs) for the enhancement of general BMD transmission. The RVNs are based on wide spreading self-organized marine vessels integrated with sensors and multi-domain communication technologies. Data forwards are along opportunistic vessel-to-vessel (V2V) connections, formed when vessels move into the communication ranges of each other. The real-time network topology is highly dynamic and resilient. The distinctive advantages of RVN based

*This work was supported in part by the National Key R&D Program of China under Grant. 2020YFB1707700, the National Natural Science Foundation of China Grant No. 61902367, and China Postdoctoral Science Foundation No. 2019M652475. *Corresponding author: Chao Liu*

Ruobing Jiang and Chao Liu are with the School of Computer Science and Technology, Ocean University of China, QingDao, 266100, China (e-mail: {jrb, liuchao}@ouc.edu.cn).

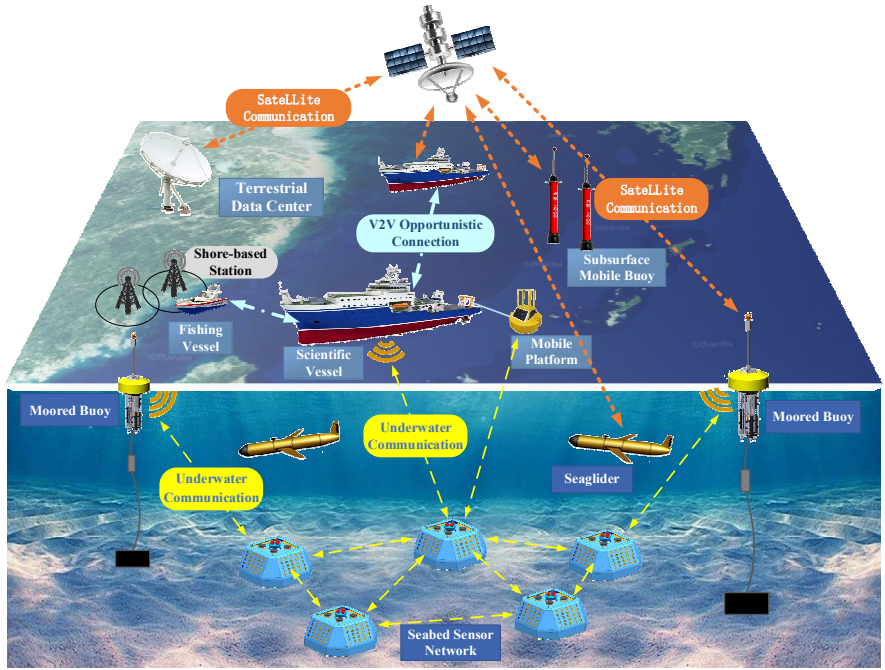

**Fig. 1 Multi-dimensional Big Marine Data (BMD) and existing communication technologies**

communication include low transmission cost, sufficient coverage, and cross-domain transmission capability, which perfectly bridge the gap between the requirements of BMD transmission and existing marine communication technologies. To predict and manage the opportunistic V2V connections among vessels for optimal data routing, we apply Social Network Analysis (SNA) on large-scale datasets of real vessel GPS trajectories to explore essential social characteristics of RVNs. Implicit social communities are detected based on vessel familiarity abstracted from historical trajectories. Social community-based routing (CBR) is then proposed to effectively utilize opportunistic connections, where marine data are selectively forwarded to vessel relays being more familiar with the targeted data receiver.

**Performance Evaluation.** The proposed CBR is evaluated with real GPS datasets of fishing vessel trajectories, comparing to typical routing algorithms. The GPS datasets are collected from 12,209 fishing vessels with diverse fishing types within the area of 121°E - 128°E, 26°N - 34°N in East China Sea. The GPS record sequences are from January 1, 2015 to December 31, 2018 with a general interval of around 3 min. Transmission performance on average delivery ratio and transmission cost are evaluated and the proposed CBR achieves largely reduced transmission cost with comparable delivery ratio.

The rest of this article is organized as follows. The following section introduces the fundamentals of social network analysis and its applications. We then model resilient vessel networks, apply SNA and propose social community-based routing in Section III. Section IV evaluates the routing performance and Section V presents the related work. Future directions and conclusion are finally given.

## II. BASICS ON SOCIAL NETWORK ANALYSIS

Social Network Analysis (SNA) [2] studies social interactions to reveal inherent patterns of individual and group behaviors. SNA is an interdisciplinary academic field involving sociology, economics, psychology, graph theory, data science, statistics, and communication. Originate from sociology, early SNA efforts include several important revelations on social phenomena, such as small world phenomenon or six-degree of separation, weak tie theory, the concept of structural hole, and Dunbar's number of human cognitive limit. Recent developments on SNA turn to quantitative network analysis based on a series of hierarchical evaluation metrics.

In this section, quantitative measurements and key issues at the node level, link level, and subnetwork level are respectively discussed. We then introduce the applications of these metrics in solving important issues of online social networks and cyber-physical networks, e.g., influential user identification, user recommendation, information cascade analysis, and social aware routing.

### A. Centrality and Node Classification

Since the main components of social networks are nodes and the associated interactions among them, the primary efforts of SNA are to understand the characteristics of network nodes by measuring and differentiating their importance. Node importance are commonly measured by centrality applying different quantitative models, e.g., Degree Centrality, Closeness Centrality, Betweenness Centrality, Katz Centrality, and Eigenvector Centrality.

*Degree Centrality* measures the number of links for each node to quantify nodes' neighborhood scale. When the interactions in a social network are designed to be directed, in-degree centrality and out-degree centrality can be considered separately. However, since node degree only involves local situation, degree centrality cannot accurately represent the global importance of a node.

*Closeness Centrality* measures the reciprocal of the average distance of the shortest paths from a node to all the other nodes in the network. The higher this metric, the closer the node is to other nodes, and thus the node is considered to be more central.

*Betweenness Centrality* is also based on shortest paths while measuring the proportion of the shortest paths that passing through a node without ending at or starting from it. This metric represents the degree to which a node bridges other node pairs in the whole network. The betweenness also has various extensions by replacing pairwise shortest paths with other paths such as random-walk betweenness, network flow betweenness and routing betweenness.

*Katz Centrality* measures node influence by taking all paths passing through the node into consideration so that all network links are involved instead of only the ones along pairwise shortest paths. This metric exhibits extremely high degree of computational complexity.

*Eigenvector Centrality* measures a node by the measurements of its neighbors. By considering both the number of links and the measurements of connected nodes, this metric requires iterative calculations to converge. Similar metrics include PageRank and its variants such as TunkRank and TwitterRank.

### B. Homophily and Link Prediction

At the link level, an important and promising task is to predict link formation. The basic idea to determine the probability of new link formation between two nodes without direct connection is to measure the structural homophily or attribute similarity of the two nodes.

For the class of link prediction methods measuring structural homophily, the assumption is that the more similar the network structure around the two nodes, the more likely they are to form a new link. Measurements such as *Common Neighbors*, *Jaccard Coefficient*, and *Adamic/Adar* are based on the number or degrees of the common adjacent nodes of the two nodes. *Katz Measure* considers the weighted sum of the lengths of the existing paths between the two nodes. *Random Walk Measure* calculates the stable arrival probability from the one node to the other by random walk.

For the class of link prediction methods measuring attribute similarity, the assumption is that two similar nodes tend to form a link. Optional attributes can be Degree, Euclidean Distance and even the similarity degree of their common neighbors

### C. Modularity and Community Detection

When nodes can be differentiated or divided by centrality or homophily, the social network will naturally exhibit the existence of community structure, where internal nodes are more densely connected with each other than external nodes.

To detect disjoint communities without common nodes, the most popular detection/partition method is *Modularity Maximization*. This method searches over all the possible network divisions to maximize the modularity benefit based on the structural metrics such as link betweenness.

The detection of overlapping communities is more challenging since any node can belong to multiple communities. Typical detection methods include node-centric ones such as *Clique Percolation* based on k-clique detection and link-centric ones such as *Link Clustering* based on hierarchical similarity measurements.

### D. SNA on Online Social Networks

As the proliferation of virous online social services such as instant messaging applications WeChat and Skype, microblogs Twitter and Weibo, and science websites Web of Science and DBLP, plenty of online data on social communities and interactions become available and facilitate the rapid development of computational SNA to perform deep mining, understanding and prediction on online social behaviors and network evolutions.

At the node level, the centrality measurements and node classification enable the identification of influential users [3], inference of unknown user attributes, and prediction of user behavior.

At the link level, the link mining enables the inference of trust relationship, recommendation of friends or products or news, prediction of criminal associations, and improvement of marketing strategies.

At the network/subnetwork level, community detection supports the analysis and management of information cascade, effective advertisement of relevant information, identification of rumors, malicious activities, and frauds, abstraction of super-scale network at different resolutions for macro decision-making or visualization.

### E. SNA on Cyber-physical Systems

Cyber-physical systems (CPS) are formed by physical nodes interacting with each other under computationally intelligent mechanisms to perform specific functions. When the physical nodes are inherently mobile, the system is a mobile CPS[4].

Typical (mobile) CPSs are wireless sensor networks [5], smart grids, vehicular networks [6][7], resilient vessel networks (RVNs), and smartphone networks. Since CPSs are serving or controlled by human behaviors, there are increasing advantages to take social factors into account.

Node level SNA identifying influential nodes improves routing or information dissemination efficiency by assigning higher forwarding priorities to more influential modes.

Link level SNA predicts future links [8] which particularly contributes to opportunistic transmissions.

Community level SNA detects communities by grouping similar nodes or links in terms of higher inner interactions within communities. Community-aware routing [9] thus reaches target nodes faster through the members of the same community.

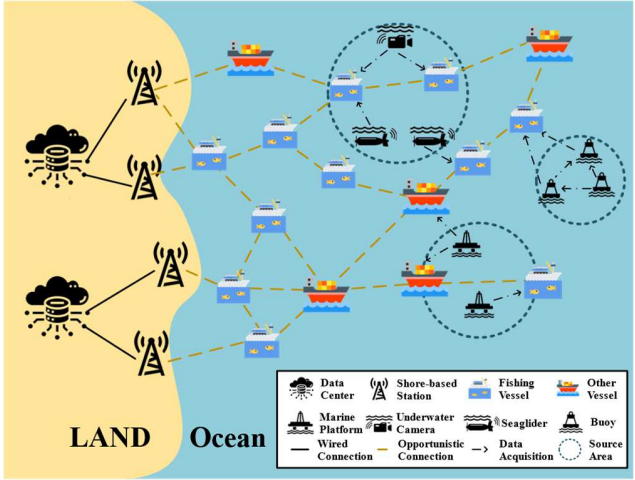

**Fig. 2 BMD transmission in Resilient Vessel Networks**

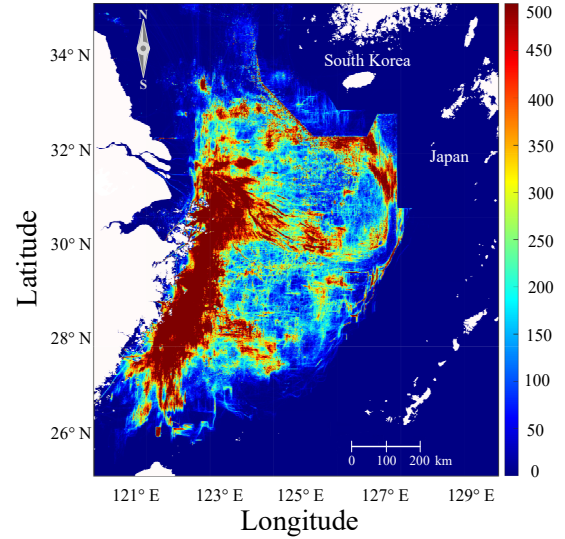

**Fig. 3 Distribution and density of fishing vessels**

## III. Socially-Aware Routing in Resilient Vessel Networks

To enhance the transmission of the big marine data with extremely huge volume, high generating velocity, multi-domain variety and heterogeneity, we propose socially-aware resilient vessel networks (RVNs) to compensate for the limitations of existing communication technologies. In this section, we first briefly introduce general BMD transmission requirements and the vessel network model. We then perform social community detection based on a novel familiarity measurement. Finally, a social community based routing (CBR) mechanism is proposed to enhance BMD transmission.

### A. Resilient Vessel Network Model

A general BMD transmission scenario based on RVNs is illustrated in **Fig. 2**. There can be virous kinds of data sources distributed underwater or over the ocean surface. We assume the underwater marine data can be sensed by the underwater sensors attached to each ocean vessel. Different marine data targeted to different data centers are transmitted in the form of packets with multi-hop opportunistic vessel-to-vessel (V2V) forwarding through shore-based wireless access stations. Each packet is tagged with identification information such as source area, target stations, and generated time stamp. An opportunistic V2V connection appears only when the two vessels enter the communication range of each other.

### B. Social Network Analysis on RVNs

We perform SNA on a real-world large-scale fishing vessel network based on real vessel trajectory data sets. The vessel trajectories are spatial-temporal positions of more than 12,000 fishing vessels from Jan. 2015 to Dec. 2018 by BeiDou Navigation Satellite System. **Fig. 3** illustrates the distribution and density of the trajectory accumulation within only one month, spreading the wide area of East China Sea within 121° E-128°E and 26°N-34°N. It is obvious that beyond as far as 100 km away from the coastline without cellular coverage, there are still non-negligible scale of vessel trajectories which can provide low-cost and high-throughput RVN based data transmission.

We perform link-level social strength analysis and network-level community detection by modeling both vessels and target stations as network nodes and historical encounters as links with encounter frequency as link weights.

We propose a novel measurement for *Node Familiarity* to better evaluate link strength, based on which vessel communities can be accurately detected to improve marine data transmission efficiency. In addition to the encounter frequency $M_{ij}$ between two vessels $i$ and $j$, we further consider the standard deviation of their encounter interval $\sigma_{ij}$, and the familiarity $F_{ij}$ between vessel $i$ and $j$ formulates as follows,

$$F_{ij} = M_{ij}^{\alpha}/\sigma_{ij}^{\beta}, \quad (1)$$

where $\alpha$ and $\beta$ are constants to adjust the influence degree between the two factors. On the one hand, higher encounter frequency between two nodes contributes to higher familiarity. On the other hand, lower standard deviation of encounter intervals reflects more stable encounters.

Based on the proposed measurement on familiarity, we apply Infomap to perform community detection and evaluate the detection performance with a small dataset of 600 same-type fishing vessels for the purpose of clarity. As shown in **Fig. 4**, the proposed familiarity based community detection achieves excellent performance where all the vessels geographically cluster into 12 obvious communities.

### C. Community-based Routing in RVNs

We propose socially aware routing to improve data forwarding efficiency by optimizing the familiarity of relays with data targets. Considering that only when vessels encounter with other, data forwarding could occur, thus the opportunistic encounters should be efficiently utilized. Only when the encountered vessel is more familiar with the data target, the data

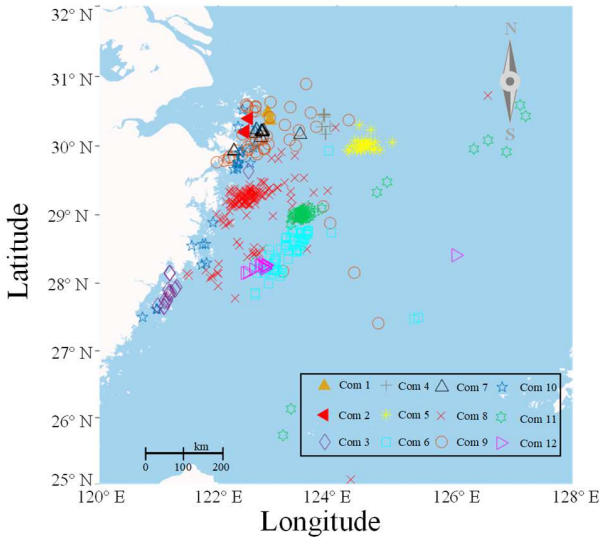

**Fig. 4 Community detection performance in RVNs**

is forwarded. Or, the encounter can be used to forward other data packets with higher benefit.

The basic idea of the proposed socially aware routing, i.e., Community-Based Routing (CBR), is to forward packets to the vessels in the same or closer community with the target stations. Two communities sharing more vessels are thought to be closer. Suppose vessel $v_0$, carrying a packet towards target station $v_d$, encounters vessel $v_i, i \neq 0$, at time $t$ and CBR makes packet forwarding decisions as follows.

1) **Community Checking**

Vessel $v_0$ first checks the community $c_i$ of $v_i$. When $v_i$ is within the same community as $v_d$ or community $c_i$ is closer to $c_d$ than $c_0$, $v_i$ is selected to be a candidate relay.

2) **Delay Minimization**

When multiple candidates exist at time $t$, $v_0$ estimates expected delivery delay through all the current candidates to derive a local optimal relay, e.g., $v_n$, by minimizing the delivery delay.

3) **Global Optimization**

$v_0$ compares the locally minimal delivery delay at time $t$ provided by $v_n$ with the globally minimal delay derived over all the historical local minimum at any time before $t$. If $v_n$ provides a global gain by offering an even shorter delay, $v_n$ is forwarded and the global shortest delivery delay is updated.

The advantage of the proposed CBR is to cut down unnecessary packet forwarding to reduce transmission cost.

## IV. TRANSMISSION PERFORMANCE EVALUATION

In this section, we evaluate the transmission performance of the proposed socially-aware community-based routing CBR through extensive simulations on large-scale dataset of real fishing vessel trajectories. We first introduce the dataset and simulation setup and then present performance metrics and typical routing algorithms for comparison. The simulation results are finally explained.

### A. Vessel Trajectory Dataset and Simulation Setup

The vessel trajectory dataset includes spatial-temporal positions of 12,209 fishing vessels of various fishing types recorded from January 1, 2015 to December 31, 2018 with BeiDou Navigation Satellite System. Vessel position records with about 3 min intervals form vessel trajectories, which spread within 121°E-128°E and 26°N-34°N of East China Sea. Note that at any given time, only a subset of fishing vessels are on sailing.

The simulations are conducted with multiple fixed areas as data sources, periodically generating data packets and uploading to vessels within the areas. Three real-world ports are set as shore-based stations collecting multi-hop relayed packets. The wireless communication range is set as 30 km. Each packet has a 2-hour lifetime.

### B. Performance Metrics and Compared Routing Algorithms

Two important transmission metrics, delivery ratio and transmission cost, are evaluated and three typical routing algorithms are implemented for comparison. The metric of delivery ratio is the average rate of successfully delivered packets to the total generated packets. The metric of transmission cost is the number of relay vessels or packet copies per delivered packet. Three routing algorithms for comparison are as follows.

- *Familiarity-based Routing (FBR)*: Different from CBR considering both node-level familiarity and community-level distance, FBR only considers node-to-node familiarity. Currently encountered vessels more familiar with the target node than previously encounter vessels will be forwarded.
- *Flooding*: Packets are forwarded to any encountered vessel. Flooding marks the upper bound of achievable delivery ratio when packet storage and transmission bandwidth are unlimited.
- *Random Walk*: Packet forwarding decisions are made with a fixed probability that can be randomly set. The probability is set as 60% in the simulations.

Note that, we assume unlimited transmission bandwidth and storage for all the compared routing algorithms to reveal the upper bound of delivery ratio by Flooding.

### C. Simulation Results

The average delivery ratio of all the routing algorithms varying with the number of vessels are shown in **Fig. 5**. For all the routing algorithms, it is obvious that more packets will be delivered with more vessels, which provides better coverage and network connectivity. The delivery ratio of the proposed CBR reaches on average 93.07% of the upper bound marked by Flooding and 11.27% higher than FBR and 82.3% higher than Random Walk. For FBR, the reason for lower delivery ratio is the lack of community-level macro navigation. Indeed, node-level familiarity maximization of FBR is easy to fall into local optimum.

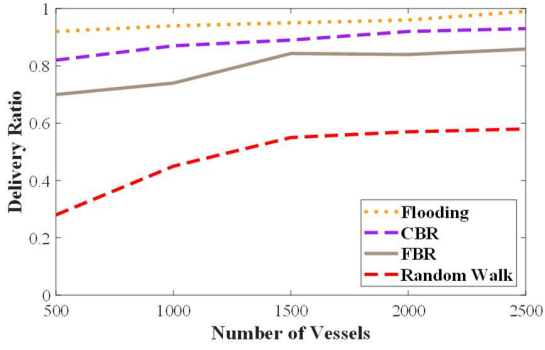

Fig. 5 Performance of Delivery Ratio

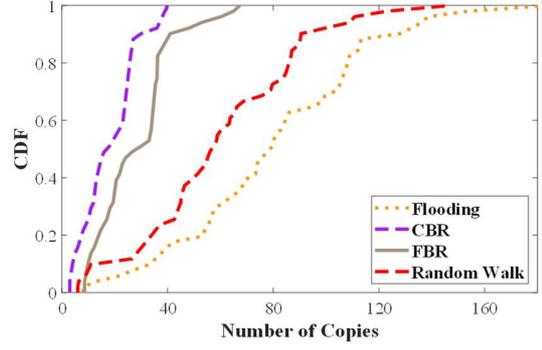

Fig. 6 Performance of Transmission Cost

The transmission cost for all the successfully delivered packets for all the compared algorithms are shown in **Fig. 6**. The proposed CBR achieves the lowest transmission cost, where 90% of the delivered packets cost less than 30 copies. FBR requires 20-40 hops, Random Walk needs 40-90 hops, and Flooding, also marking the upper bound of the transmission cost, costs the most.

The advantages of CBR over other compared algorithms are the community-based macro navigation and the hierarchical delay minimization at both intra-community level and inter-community level.

## V. RELATED WORK

We briefly classify existing communication techniques for marine data transmission into four categories. Recent advances within the four research areas are introduced in this section.

### A. Resilient Vessel Networks

Resilient vessel networks have already been used for marine data collection and transmission for the significant advantages of low-cost, high flexibility, and large-scale coverage. Existing advances are focused on ad hoc networking, integrating with other communication facilities [10] such as cellular based stations and satellites, and socially-aware transmission [9].

For ad hoc networking, typical networking and transmission techniques from terrestrial ad hoc networks like vehicular ad hoc network can be adapted to use. For example, the ideas of geographically greedy routing, dynamically hierarchical clustering, and utilizing or predicting future trajectories [11], are also applicable to vessel networks.

The multi-agent integration networking is a new trend for marine communication, which involves vessel networks, cellular based stations, satellites, and unmanned aerial vehicles (UAV), incorporating cloud computing, mobile edge computing, and deep reinforcement learning.

### B. Shore-based Cellular Communication

Shore-based cellular communication has the disadvantage of limited coverage of land-based base stations. Traditional applicable scenarios include near-coastline area, islands area, and marine platforms connected by fiber optic cable.

Recent research explores the possibilities of maritime 5G communications. The combination of 5G and unmanned aerial vehicles (UAVs) has been put forward to extend cellular coverage to remote marine area by dynamic UAV networking as relays. The integration of 5G with existing vessel navigation and monitoring system has also been proposed. The resulted intelligent communication platform is expected to facilitate vessels communicate with each other and avoid information island.

### C. Satellite Communication

Satellite communication systems provide stable, real-time, comprehensive coverage but most expensive communication services, yet requiring specific terminals for signal reception and sending.

The maritime applications of satellites including both communication and remote sensing. Typical satellite communication systems include O3b for the users without Internet access, Iridium and Globalstar providing global personal communication services. Recent micro-satellite projects [12], aiming to enable global Internet access through satellite networks, include Starlink, OneWeb, and Telesat. Recent advances in marine satellite communication also integrate UAVs to construct hybrid satellite-terrestrial maritime communication network for coverage enhancement.

For remote sensing, satellite-borne Synthetic Aperture Radar (SAR) has been utilized for worldwide maritime surveillance. The high-resolution imagery data captured by SAR platforms can be used for ship detection over broad ocean areas and in all weather conditions. Satellite-borne SAR is developed to overcome the coverage limitations of coastal systems.

### D. Underwater Communication

Underwater communication mainly includes acoustic, laser, and magnetic induction technologies. Underwater acoustic communication [13] involves limitations on limited bandwidth and data rate and significant channel loss. Thus, modulation techniques are the key research area of underwater acoustic communication. Underwater laser or light communication also suffers high absorption and severe scattering, which severely limit communication range. Optical signals yet need to be accurately pointed and cannot adapt to dynamic application scenarios. All the above communication media cannot support seamless cross-domain communication to transmit cross air-water interface. Cross-domain electromagnetic communication is more robust but attenuate dramatically in underwater environment for remote and high frequency propagation. The exploration on MIMO communication mechanism integrating multiple signals is promising.

## VI. Future Directions

Performance evaluation results demonstrate that the proposed socially-aware RVNs enhance marine data transmission. However, there are still quite a few possible directions worth being explored.

**Performance Optimization for Vessel Networks:** As previous mentioned, the performance of vessel networks heavily relies on the mining of vessel mobility patterns. In addition to social attributes utilized in this paper, other vessel attributes as future trajectory, fishing type, and points of interest can also be fully investigated. For example, if future trajectories can be accurately predicted, future vessel encounters can be achieved and managed in advance. Similarly, work type and points of interest could also be considered. Deep mining of vessel mobility patterns and the trade-off among them require further studies.

**QoS Classification and Assurance:** Real-world data transmission applications involve different priorities. For instance, ocean monitoring data for marine disaster forecasting might be more urgent than scientific data and should be transmitted within higher priority and QoS assurance. A well-established network architecture should have a set of technologies to guarantee its ability to run high-priority applications and work dependably under limited network capacity. Moreover, resilient vessel networks operate in fully distributed way, putting forward higher demands on hierarchical QoS assurance. Novel QoS mechanisms are highly nontrivial and should be well developed.

**Multi-modal Network Collaboration:** Multi-modal network collaboration can significantly improve system performance by properly resource assignments. This work only considers mobile vessels as ad-hoc nodes for system design. Multi-modal communication modes [14], such as satellite, VHF, cellular, UAV supported, are not involved. Building a proper mechanism among various network modes is still an open problem. The resource allocation issue among these conflict wireless channels [15] is worth research in the future.

**Design of Innovative Network Architecture:** Software-defined networking (SDN) is considered one of such emerging architectures with promising advantages compared to traditional networks, which is suitable for multi-application maritime resilience networks. Although SDN has attracted much attention with many benefits, unique challenges, such as link uncertainty, severe data delivery delay, and multi-modal cooperation, should be tackled.

## VII. Conclusion

In the current era of marine data proliferation, significant transmission challenges arise towards the acquisition, collection and transmission on the big marine data. To bridge the great gap between the challenges and the extremely limited transmission capability of existing transmission technologies, the new type of RVNs is proposed to provide low-cost and spatial-temporal coverage with the huge volume of existing marine vessels. Moreover, social community detection with novel measurement on node familiarity is leveraged to improve data transmission performance of RVNs by exploring the implicit social relationships among vessels. Specifically, vessel familiarity and social communities are detected.

BIOGRAPHIES

Ruobing Jiang (jrb@ouc.edu.cn) received her Ph.D. degree in Computer Science and Technology in Shanghai Jiao Tong University in 2017. She is currently an assistant professor in the Department of Computer Science and Technology in Ocean University of China. Her main research interests include mobile computing, acoustic sensing, and network security. She has authored or co-authored more than 30 technical papers, including INFOCOM 2021, 2022, TMC 2018, 2017, TPDS 2015, 2014, TBD 2021, SANER 2021, and Computers & Security 2020.

Chao Liu (liuchao@ouc.edu.cn) received the PhD degree from the Ocean University of China, Qingdao, China, in 2016. He is currently an associate professor in the Department of Computer Science and Technology, Ocean University of China, Qingdao, China. His main research interests include acoustic sensing, mobile computing, and wireless sensor networks. He has authored or coauthored more than 40 papers in international journals and conference proceedings, such as the IEEE INFOCOM, the IEEE TRANSACTIONS ON IMAGE PROCESSING, the IEEE TRANSACTIONS ON INDUSTRIAL INFORMATICS, and the IEEE TRANSACTIONS ON CIRCUITS AND SYSTEMS FOR VIDEO TECHNOLOGY. He is a member of the ACM and IEEE.